\documentclass[twocolumn]{revtex4}[12pt]
\usepackage{graphicx}
\begin{document}
\title{Modulational instability of quantum electron-acoustic waves and associated envelope solitons in a degenerate quantum plasma}
\author{Foisal B. T. Siddiki$^{a)}$, A. A. Mamun$^{b)}$, and M. R. Amin$^{c)}$}
\address{$^{a)}$Department of Electrical and Electronic Engineering, University of Dhaka, Ramna, Dhaka 1000, Bangladesh\\
$^{b)}$Department of Physics, Jahangirnagar  University, Savar, Dhaka 1342, Bangladesh\\
$^{c)}$Department of Electronics and Communications Engineering, East West University, Aftabnagar, Dhaka 1212, Bangladesh}
\date{\today}
\begin{abstract}
The basic features of linear and nonlinear quantum electron-acoustic (QEA) waves in a degenerate quantum
plasma (containing non-relativistically degenerate electrons,
superthermal or $\kappa$-distributed electrons, and stationary
ions) are theoretically investigated.  The nonlinear
Sch\"{o}dinger (NLS) equation is derived by employing the
reductive perturbation method. The stationary solitonic solution of the NLS equation are  obtained, and examined analytically as well as numerically to identify the basic features of the QEA envelope solitons.
It has been found that the effects of the degeneracy and
exchange/Bohm potentials of cold electrons, and superthermality of hot electrons significantly modify the
basic properties of linear and nonlinear QEA waves.  It is observed that the QEA waves are
modulationally unstable for $k<k_c$, where $k_c$ is the maximum (critical) value of the QEA wave number $k$
below which the QEA waves are modulationally unstable), and that for  $k<k_c$  the solution of the NLS equation gives rise to  the bright envelope solitons, which are found to be localized in both spatial ($\xi$) and time ($\tau$) axes.  It is  also observed that as the spectral index $\kappa$ is increased, the critical value of the wave number  (amplitude of the  QEA envelope bright solitons)  decreases (increases).  The implications of our results
should be useful in understanding the localized electrostatic
perturbation in  solid density plasma produced by irradiating metals
by intense laser, semiconductor devices, microelectronics, etc.
\end{abstract}
\maketitle
\section{Introduction}
The signature of electron-acoustic (EA) waves was first observed in
the laboratory experiment of Derfler  and Simonen \cite{DS1969}. This
led Watanabe and Taniuti \cite{WT1977} to consider a plasma
containing  electron species of two distinct
temperatures and  ions, and led to predict theoretically the existence of the EA
waves \cite{YS1983} in which the restoring force (inertia)  is provided  by hot electron-temperatre (cold electron mass). The EA wave frequency ($\omega$), in fact, satisfies the condition $\omega_{pi}\ll \omega\le \omega_{pc}$, where $\omega_{pi}$ ($\omega_{pc}$) is the ion (cold electron) plasma frequency.
This means that
 in the EA waves ions are reasonably assumed to be stationary, and  to maintain only the neutralizing
background.  The dispersion relation for the
long-wavelength (in comparision with the hot electron Debye length) EA waves
is \cite{YS1983} $\omega\simeq
kC_e$, where $k$ is the wave number, and
$C_e=(n_{c0}T_h/n_{h0}m_e)^{1/2}$  [where $n_{c0}$ ($n_{h0}$) being the unperturbed
cold (hot) electron number density, $T_h$ being the hot electron temperature in
units of the Boltzmann constant, $m_e$ being the cold electron mass]  is the electron-acoustic speed.
The long wavelength EA waves are also detected in space plasma environments
\cite{TG1984,M-etal2001,SL2001}.   The conditions for the existence
of the linear EA waves and their dispersion properties are now
well-understood from both theoretical  \cite{WT1977,YS1983} and experimental  \cite{DS1969,HT1972}
points of view.  The basic properties of the nonlinear EA waves, particularly  EA solitons  in electron-ion plasmas have been investigated by several authors
\cite{MH1990,Mace-1991,MS2002,ref39}.

The nonlinear structures in degenerate plasmas  have also
received  a renewed interest in understanding the localized
electrostatic disturbances not only in astrophysical environments
(such as neutron stars, white dwarfs, magnetars, etc.
\cite{Mamun2010a,Mamun2010b,El-Taibany2012,Hossen2015}), but also
in laboratory devices (viz.  solid density plasma produced by
irradiating metals by intense laser, semiconductor devices,
microelectronics, carbon nanotubes, quantum dots, and quantum
wells, etc. \cite{Jung2001,Ang2003,Killian2006,Shah2012}). Recent
investigations \cite{ref8,ref9,ref10,ref11} based on quantum
hydrodynamic (QHD) model\cite{Brodin2007,ref12,ref13} show a
number of significant differences in nonlinear features of quantum
plasmas from those in classical electron-ion plasmas. The QHD
model is a useful approximation to study the short-scale
nonlinear structures in dense (degenerate)  quantum plasmas
\cite{ref12,ref13}, where the effects of degenerate (instead of
thermal) pressure , exchange correlation potential, and Bohm
potential  can be included.

Recently, Zhenni et. al. \cite{ref45} have studied EA solitary waves  or  shortly EA solitons in magnetized quantum plasma with relativistic electrons, while  Chandra and Ghosh \cite{ref46} have studied the modulational instability of the  EA waves in relativistically degenerate quantum plasmas. However, they have not considered the exchange correlation  and  Bohm potentials  in their investigation.
Therefore, in our present work, we investigate linear and nonlinear propagation of the quantum EA (QEA) waves to include the effects of  superthermality \cite{ref47} of hot electron component, and  quantum effects due to the degenerate particle pressure, exchange correlation and  Bohm potentials of cold electron component.  We also study the amplitude modulation of the slow evolution of the QEA envelope solitions (QEAESs) by deriving a nonlinear Schr\"{o}dinger  (NLS) equation by taking these effects into account.

The manuscript is organized as follows. The basic equations governing the plasma system under consideration are provided in  Sec. II.  The  NLS equation for the nonlinear propagation of the EA Waves  is derived by applying the reductive perturbation technique, and  their  linear as well as nonlinear properties are examined in Sec. III.  A brief discussion is  presented in Sec. IV.
\section{Governing Equations}
We consider a three-component plasma system  containing cold quantum electron fluid with Fermi energy
$E_F$ \cite{ref12,ref13},  inertialess, superthermal  \cite{MH1990,Mace-1991} or  hot electron component,
and uniformly distributed stationary ions \cite{MS2002}.  Thus, at equilibrium we have $n_{c0}+n_{h0}=n_{i0}$, where $n_{s0}$ is the number density of plasma species $s$ ($s=c$ for cold electron species, $s=h$ for hot electron species, and $s=i$ for stationary ion species).  The dynamics of the QEA waves in such a three-component quantum plasma system is governed by the following set of QHD equations \cite{ref12,ref13,ref4,ref48,ref51}:
\begin{eqnarray}
&&\frac{\partial n_c}{\partial t}+\frac{\partial}{\partial x}\left(n_cv_c\right) =0,
\label{beq1}\\
&&\frac{\partial v_c}{\partial t} + v_c\frac{\partial v_c}{\partial x} = -\frac{e}{m_e}\frac{\partial\phi}{\partial x}
-  \frac{\partial V_{DB}}{\partial x}-\frac{\partial V_{xc}}{\partial x},
\label{beq2}\\
&&\frac{\partial^2\phi}{\partial x^2} = 4\pi e\left(n_c+n_h-n_{i0}\right),
\label{beq3}
\end{eqnarray}
where $n_c$ ($v_c$) is  the number density  (fluid speed) of the cold electron species; $\phi$ is the electrostatic wave potential;  $-e$  ($m_e$) is the  electron charge (mass); $ x$ ($t$) is the spatial (time) variable; $V_{DB}=P_c/m_en_c+V_B$,  in which $P_c$ is the non-relativistically degenerate cold electron pressure, and is given by \cite{ref4}
\begin{eqnarray}
P_c = \frac{\hbar^2\pi^{2/3}}{5m_e}n_c^{5/3},
\label{beq4}
\end{eqnarray}
with $\hbar$ being the Planck constant divided  by $2\pi$; $V_B$ is the Bohm potential,  and is given by
\cite{ref8,ref48}
\begin{eqnarray}
V_B = \frac{\hbar^2}{2m_e}
\left(\frac{1}{\sqrt{n_c}}\frac{\partial^2 \sqrt{n_c}}{\partial x^2}\right),
\label{beq9}
\end{eqnarray}
which is due to  the tunneling effect of the cold electrons; and $V_{xc}$ is the exchange-correlation potential,
and is  given by \cite{ref48,ref51}
\begin{eqnarray}
V_{xc} =-0.985e^2n_c^{1/3}\left[1+\frac{0.624}{a_Bn_c^{1/3}}\ln{\left(1+a_Bn_c^{1/3}\right)}\right],
\label{beq10}
\end{eqnarray}
with $a_B=18.37\hbar^2/m_pe^2$.
We note that the exchange-correlation potential can be separated  into two terms, namely the Hartree term due to the electrostatic potential of the total cold electron number density and the cold electron exchange-correlation potential  term \cite{ref48,ref51}.

The hot electron species is assumed to be superthermal ($\kappa$ distributed). Thus, the number density ($n_h$) of the hot electron species is given by \cite{MH1990,Mace-1991,ref47,ref39}
\begin{eqnarray}
n_h = n_{h0}\left[1-\frac{e\phi}{k_BT_h\left(\kappa-\frac{3}{2}\right)}\right]^{-\kappa+1/2},
\label{beq11}
\end{eqnarray}
where $T_h$  is hot electron temperature, and $\kappa$ is the spectral index measuring the deviation from the thermal equilibrium, and its value is $\kappa>3/2$ for superthermal electrons \cite{MH1990,Mace-1991,ref39}.

We now normalize all the variables as follows: $X=x/\lambda_D$,  $T=t\omega_{pc}$,  $V=v_c/C_e$,
$\Phi=e\phi/E_F$, $N=n_c/n_{c0}$, where $C_e=\left(E_F/m_e\right)^{1/2}$,
$\omega_{pc}=(4\pi e^2n_{c0}/m_e)^{1/2}$,
$\lambda_D=C_e/\omega_{pc}$, and  $E_F=\hbar^2(3\pi^2 n_{c0})^{2/3}/2m_e$.   Thus,  (\ref{beq1})-(\ref{beq3})  can be written in the normalized form as
\begin{eqnarray}
&&\frac{\partial N}{\partial T}+V\frac{\partial N}{\partial X}+N\frac{\partial V}{\partial X}=0,
\label{beq12}\\
&&\frac{\partial V}{\partial T} + V\frac{\partial V}{\partial X} = \alpha\frac{\partial\Phi}{\partial X}
- \frac{\partial \Psi_{DB}}{\partial X}+ \mu N^{-1}\frac{\partial^3 N}{\partial X^3},
\label{beq13}\\
&&\frac{\partial^2\Phi}{\partial X^2} -\delta\Phi-\nu\Phi^2 =N-1,
\label{beq14}
\end{eqnarray}
where $\Psi_{DB}=3(\sigma N^{2/3}+ 2\beta N^{1/3})/2$,
$\sigma=\hbar^2(\pi n_{c0})^{2/3}/m_e^2C_e^2$,
$\beta= (0.33e^2 n_{c0}^{1/3}/{m_eC_e^2})
[1+0.625/(1+18.37a_Bn_{pc0}^{1/3})]$,
$\mu=(\hbar\omega_{pc}/2m_eC_e^2)^2$,
$\delta=\alpha E_F(\kappa+1/2)/k_BT_h(\kappa-3/2)$,
$\nu =\alpha\delta E_F(\kappa+1/2)/k_BT_h(\kappa-3/2)$, and
$\alpha=n_{h0}/n_{c0}$.
\section {Nonlinear Schr\"{o}dinger Equation}
 To derive the NLS equation for slow evolution of the QEA waves by the  reductive perturbation method
\cite{Washimi1966}, we first introduce the stretched coordinates:
 \begin{equation}
\left.
\begin{array}{l}
\xi=\epsilon (X-v_0 T),\\
\tau=\epsilon^2T,
\end{array}
\right\}
\label{str1}
\end{equation}
and expand the dependent variables $N$, $V$, and $\Phi$:
\begin{eqnarray}
&&N=1+\sum_{n=1}^{\infty} \epsilon^n\sum_{l=-\infty}^\infty N_{l}^{(n)}(\xi,\tau)\: e^{il(kX-\omega T)},
\label{exp1}\\
&&V=\sum_{n=1}^{\infty}\epsilon^n\sum_{l=-\infty}^\infty V_{l}^{(n)}(\xi,\tau)\: e^{il(kX-\omega T)},
\label{exp2}\\
&&\Phi=\sum_{n=1}^{\infty} \epsilon^n\sum_{l=-\infty}^\infty \Phi_l^{(n)}(\xi,\tau)\: e^{il(kX-\omega T)},
\label{exp3}
\end{eqnarray}
where $v_0$ is the group velocity of the  QEA waves (to be determined later),   $\epsilon$ is an expansion parameter ($0<\epsilon<1$), $\omega$  ($k$) is the angular frequency  (wave number) of the carrier QEA waves. The quantities $N_l^{(n)}(\xi,\tau)$, $V_l^{(n)}(\xi,\tau)$, and $\Phi_l^{(n)}(\xi,\tau)$ are the $l$-th harmonic of the $n$-th order slowly varying dependent variables, and these satisfy the reality condition $A_l^{(n)}\equiv A_l^{(n)*}$, in which  $*$ denotes the complex conjugate of the quantity involved.

Now, substituting  (\ref{str1})- (\ref{exp3}) into   (\ref{beq12})-(\ref{beq14}),  and performing few steps of straight forward mathematics, we can obtain the 1st  harmonic of the 1st  order ($l=1$ and $n=1$) reduced equations, which allow us  to express the linear dispersion for the QEA waves as
\begin{eqnarray}
\omega^2 = k^2 \left(\sigma+\beta+k^2\mu+\frac{\alpha}{k^2+\delta}\right).
\label{disp}
\end{eqnarray}
We note here that the parameters $\sigma$, $\beta$, and $\mu$ account for the quantum effects due to the
degenerate particle pressure, particle exchange-correlation potential, and the Bohm potential, respectively, on
the linear dispersion relation for the QEA waves.  So, $k^2\sigma=T_{DP}$,  $k^2\beta=T_{XP}$, and
$k^4\mu=T_{BP}$ represent the quantum effects due to the degenerate particle pressure, particle exchange-correlation potential, and the Bohm potential, respectively.
We have  shown how these quantum effects (represented by  $T_{DP}$,
$T_{XP}$, and $T_{BP}$)  vary with the carrier QEA wavenumber $k$.

\begin{figure} [htp]
\centerline{\includegraphics[width=8cm]{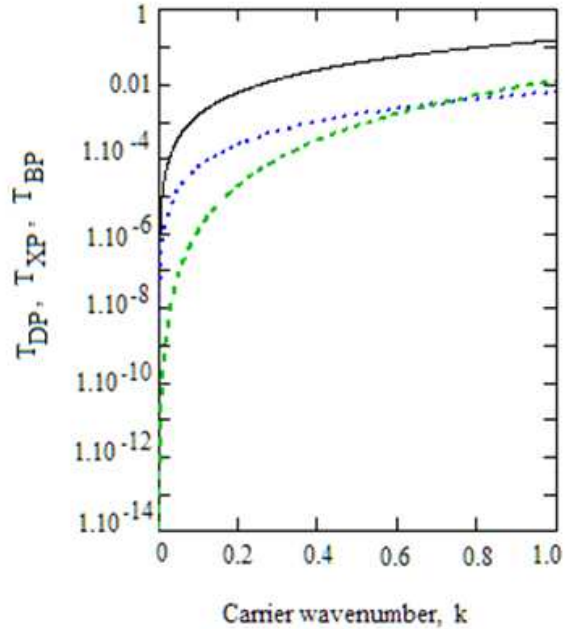}}
 \caption{The variation of the quantum terms $T_{DP}$, $T_{XP}$, and $T_{BP}$ with the QEA wavenumber $k$ for
$n_{c0}=10^{28}$ cm$^{-3}$, $\kappa=1.6$, and $\alpha=0.8$. The solid curve is for $T_{DP}$, the dotted curve is for $T_{XP}$, and the dashed curve is for $T_{BP}$.}
\label{f1}
\end{figure}

\begin{figure}[htp]
\centerline{\includegraphics[width=8cm]{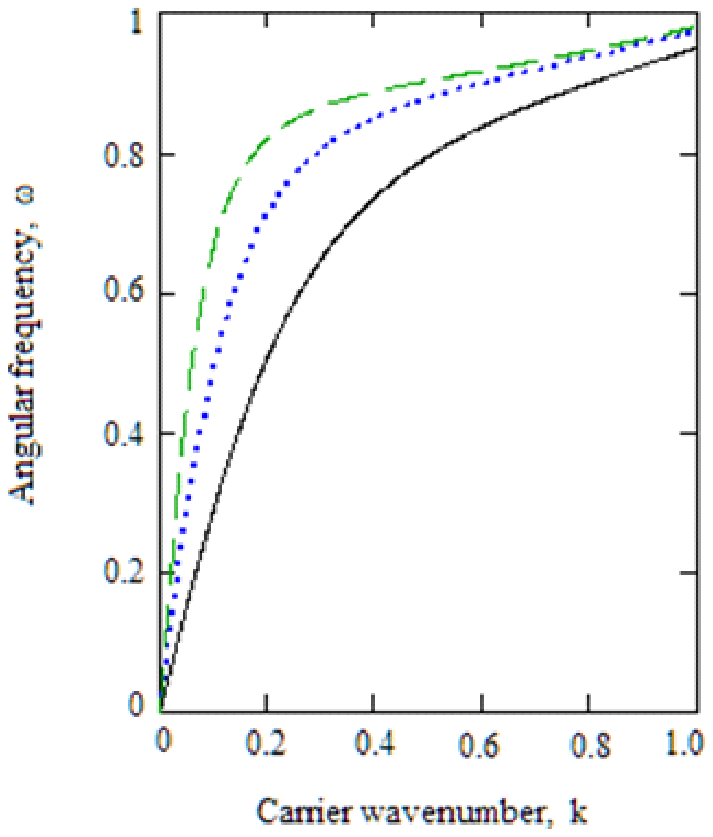}}
\caption{The dispersion ($\omega$ vs. $k$) curves of the QEA waves  for $n_{c0}=10^{28}$ cm$^{-3}$,
$\alpha=0.8$,  $\kappa=1.6$ (solid curve), $k=2$ (dotted curve), and $k=50$ (dashed curve).}
\label{f2}
\end{figure}

\begin{figure}[htp]
\centerline{\includegraphics[width=8cm]{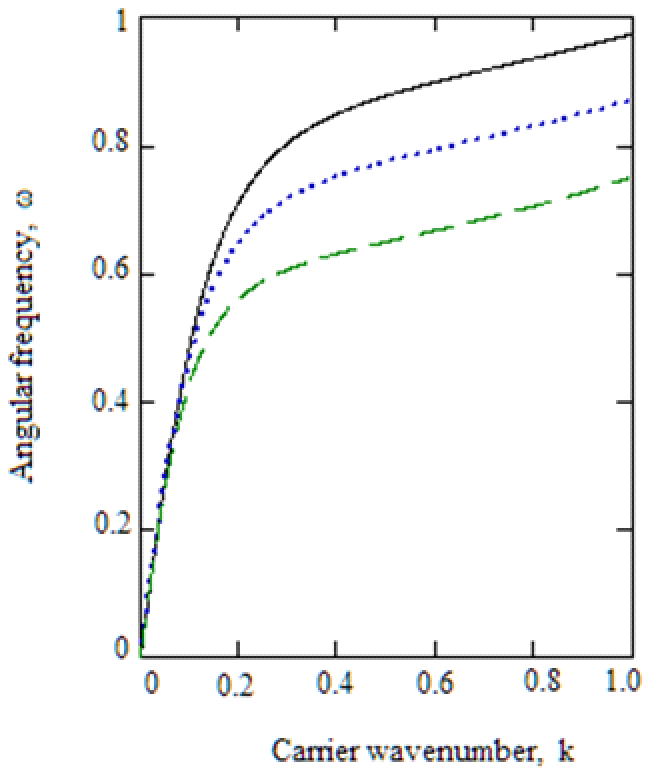}}
\caption{The dispersion ($\omega$ vs. $k$) curves of the QEA waves  for $n_{c0}=10^{28}$ cm$^{-3}$,
$\kappa=2$,  $\alpha=0.4$ (solid curve),  $\alpha=0.6$ (dotted curve), and $\alpha=0.8$ (dashed curve).}
\label{f3}
\end{figure}

This is displayed in figure \ref{f1}, where the solid (dotted) curve shows how the effect of electron degenerate pressure (particle exchange potential) varies with $k$, and the dashed curve show how the effect of Bohm potential varies with $k$.   It is observed from figure \ref{f1} that the effect of the electron degenerate pressure is  more  significant than that of both exchange-correlation  and Bohm potentials.  It is further observed from figure \ref{f1} that the effect of the exchange-correlation (Bohm) potential is more significant for the smaller (larger) values of the carrier wavenumber $k$. We have graphically shown the effects of superthermality (represented by spectral index
$\kappa$) and number density  of hot electrons  (represented by the parameter $\alpha$) on
the dispersion ($\omega$ vs. $k$) curves.  These are depicted in  figures \ref{f2} and  \ref{f3}. They indicate that as  $\kappa$ ($\alpha$) increases,  the group velocity $v_0$  increases for lower (higher) values of
$\kappa$ ($\alpha$) , and becomes very sharp at the low value ranges of $\kappa$ and $\alpha$.

It is obvious from figures  \ref{f2} and  \ref{f3} that  for long wavelength limit (which corresponds to a very low
$k$-value range) the angular frequency $\omega$ linearly increases with $k$,  and for short wavelength limit (which corresponds to a very high $k$-value range) it is independent of $k$  (saturated region). This is usual dispersion properties of any kind of acoustic-type of waves. It is observed from figure \ref{f2}  (figure \ref{f3}) that
as we  increase $\kappa$ ($\alpha$),  the $\omega$ vs.  $k$  curve is shifted up (down) to  $\omega-$ axis, and the saturation region is reached  for higher values of  $\kappa$ and $\alpha$.

\begin{figure}[htp]
\centerline{\includegraphics[width=8cm]{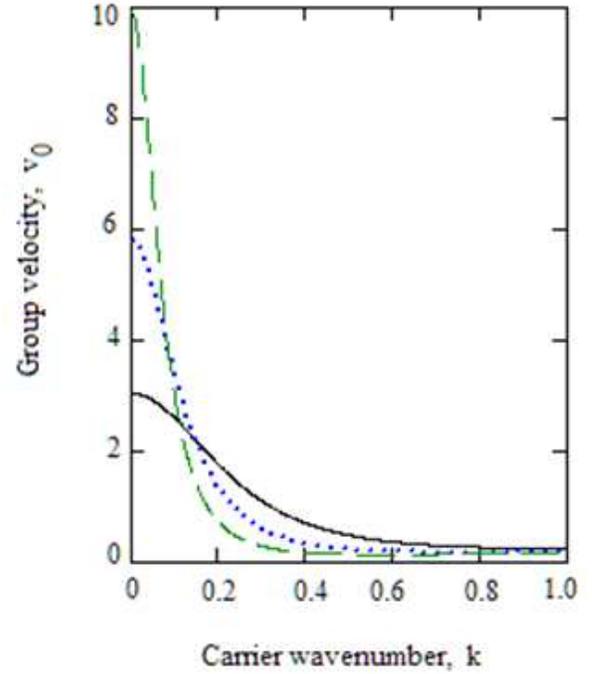}}
\caption{The variation of the QEA wave group velocity $v_0$ with the QEA wave number $k$ for $n_{c0}=10^{28}$ cm$^{-3}$,  $\alpha=0.8$,  $\kappa=1.6$ (solid curve), $\kappa=2$ (curve), and $\kappa=50$ (dashed curve).}
\label{f4}
\end{figure}

\begin{figure}[htp]
\centerline{\includegraphics[width=8cm]{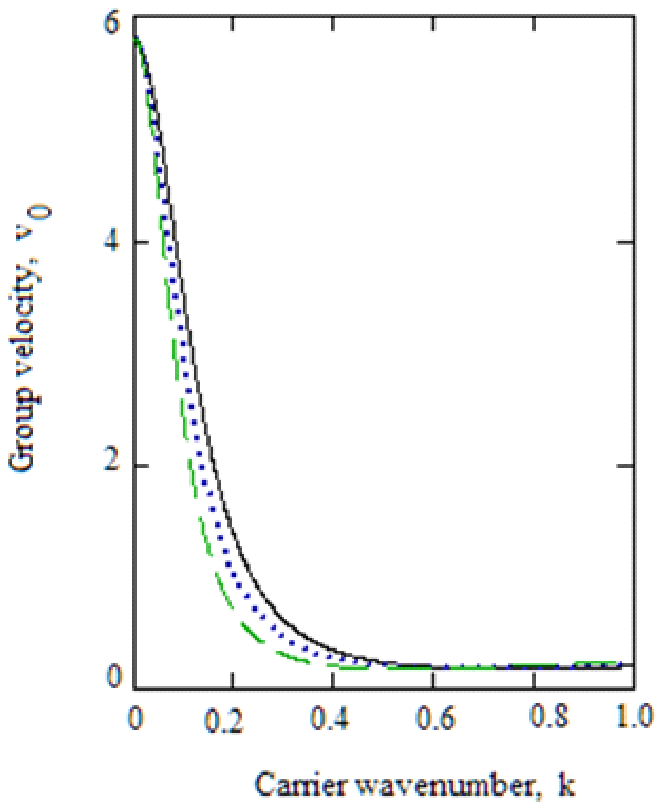}}
\caption{The variation of the QEA wave group velocity $v_0$ with the QEA wavenumber $k$ for $n_{c0}=10^{28}$ cm$^{-3}$ and  $\kappa=2$, $\alpha=0.4$ (solid curve), $\alpha=0.6$ (dotted curve), and $\alpha =0.8$ (dashed curve).}
\label{f5}
\end{figure}

Now, following the same procedure, from the first harmonic of the second order quantities ($n=2$ and $l=1$), and from (\ref{disp}), we can express  $v_0$  as
\begin{eqnarray}
v_0 = \left(\frac{1}{a_1\omega+a_2k}\right)\left[\omega a_2-kb_1+ 2k\left(\omega^2-k^2b_0\right)\right],
\label{group}
\end{eqnarray}
where
$a_1=-k^2\alpha/\left(\omega^2-k^2b_0\right)$, $a_2=\omega a_1/k$,  $b_0=\sigma+\beta+k^2\mu$, and $b_1=\alpha+a_1b_0$.  It should be mentioned here that in our present investigation we are interested in the low-frequency, long wavelength  QEA waves. We have graphically shown the effects of superthermality (represented by the spectral index $\kappa$) and hot electron number density (represented by the parameter $\alpha$) on $v_0$ vs. $k$ curves.
The results are depicted in figures \ref{f4}  and \ref{f5}.
Now, from the 2nd harmonic of the second order ($l=2$ and $n=2$) reduced equations, we can express  $\Phi_2^{(2)}$  in terms of  $\Phi_1^{(1)}\Phi_1^{(1)}$, which  arises  from the nonlinear self-interaction. Similarly,  from  the zeroth harmonic of the third order ($l=0$ and $n=3$) reduced equations, we can express $\Phi_0^{(2)}$  in terms of $\left|\Phi_1^{(1)}\right|^2$.  We finally substitute  $\Phi_2^{(2)}$ and $\Phi_0^{(2)}$ into the 1st harmonic  $l=1$ of 3rd order ($n=3$) reduced equations to obtain the following NLS equation for the slow evolution of the QEA waves in the form
\begin{eqnarray}
i\frac{\partial a}{\partial\tau}+P\frac{\partial^2a}{\partial\xi^2} +Q|a|^2 a = 0,
\label{NLS}
\end{eqnarray}
where $a\equiv \Phi_1^{(1)}$,  and  the dispersion and nonlinear coefficients $P$ and $Q$ are
\begin{eqnarray}
&&P = \left[-\frac{k^2\alpha}{a_1}+\omega f_1+kf_2\right]\left[\omega a_1+ka_2\right]^{-1},
\label{NLS1}\\
&&Q = \left[\frac{2k^2\alpha\nu}{a_1}f_0+\omega g_1+kg_2\right]\left[\omega a_1+ka_2\right]^{-1},
\label{NLS2}
\end{eqnarray}
in which $f_0$, $f_1$, $f_2$, $g_1$, and $g_2$ are listed in  the Appendix.
\begin{figure}[htp]
\centerline{\includegraphics[width=8cm]{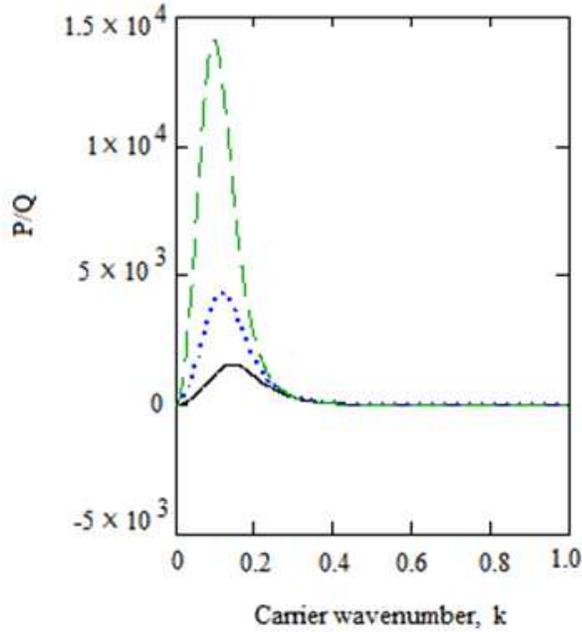}} \caption{The
variation of the ratio $P/Q$  with $k$ for $n_{c0}=10^{28}$
cm$^{-3}$,   $\alpha=0.8$,  $\kappa=1.7$ (sold curve),
$\kappa=1.8$ (dotted curve), and  $\kappa=2.0$ (dashed curve).}
\label{f6}
\end{figure}
\begin{figure}[htp]
\centerline{\includegraphics[width=8cm]{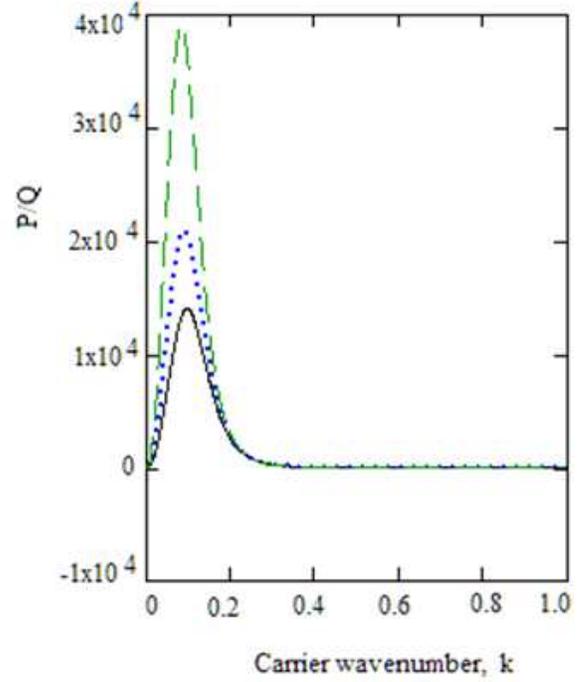}}
\caption{The variation of the ratio $P/Q$ with $k$ for  $n_{c0}=10^{28}$ cm$^{-3}$,  $\kappa=2$,  $\alpha=0.4$ (solid curve), $\alpha=0.6$ (dotted curve), and $\alpha=0.8$ (dashed curve).}
\label{f7}
\end{figure}
\begin{figure}[htp]
\centerline{\includegraphics[width=8cm]{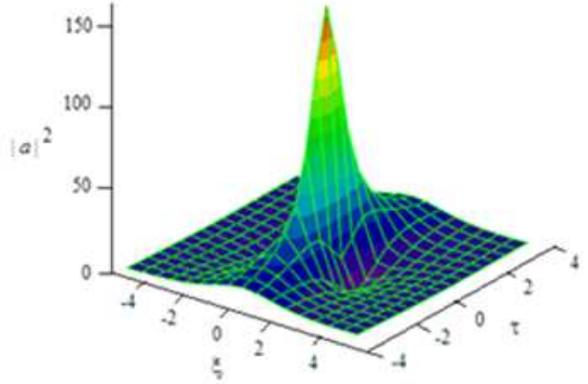}}
\caption{The time dependent envelope solitonic profiles of  $|a|^2$ for $n_{c0}=10^{28}$ cm$^{-3}$, $\alpha=0.6$, and $\kappa=1.8$.}
\label{f8}
\end{figure}
\begin{figure}[htp]
\centerline{\includegraphics[width=8cm]{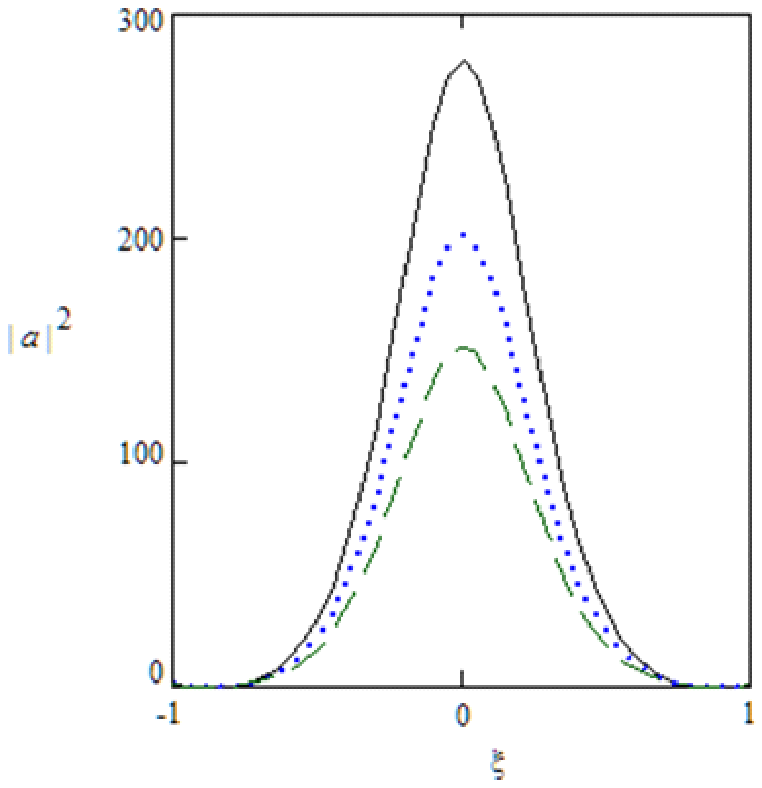}}
\caption{The envelope solitonic profiles of $|a|^2$  for $\tau=0$,
 $n_{c0}=10^{28}$ cm$^{-3}$, $\alpha=0.8$,  $\kappa=1.6$ (solid curve),
$\kappa=1.8$ (dotted curve), and $\kappa=2$ (dashed curve).}
\label{f9}
\end{figure}
\begin{figure}[htp]
\centerline{\includegraphics[width=8cm]{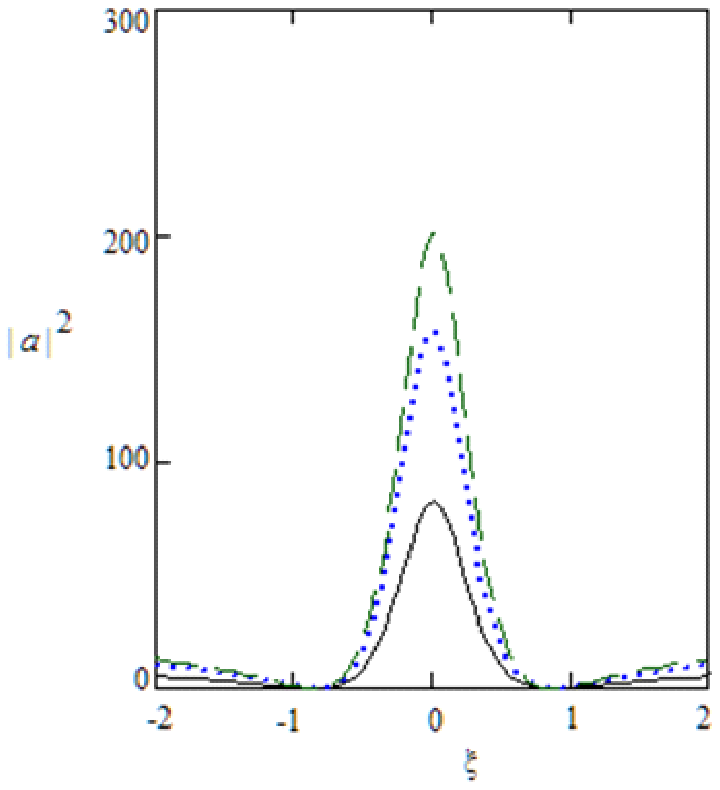}}
\caption{The envelope solitonic profiles of  $|a|^2$ for $\tau=0$, $n_{c0}=10^{28}$ cm$^{-3}$, $\kappa=2$,
$\alpha=0.4$ (solid curve), $\alpha=0.6$ (dotted curve), and $\alpha=0.8$ (dashed curve).}
\label{f10}
\end{figure}
The signs of $P/Q$ determine whether the slowly varying wave amplitude is modulationally stable or not. If  $P/Q<0$, the wave amplitude is modulationally stable, and the corresponding solution of the NLS equation is called a dark soliton \cite{ref56}. On the other hand, if $P/Q>0$,  the wave amplitude becomes modulationally unstable, and the solution of the NLS equation in this case is called a bright soliton \cite{ref56}.  We have graphically shown how $P/Q$ varies with $k$ for different values of $\kappa$ and $\alpha$. These are dipicted in figures \ref{f6} and  \ref{f7}.
It is observed from figures \ref{f6} and \ref{f7} that  $P/Q$ is positive for lower values of the carrier wavenumber $K$,  and  it ($P/Q$) changes sign from positive to negative after a certain carrier wavenumber $k=k_c$, known as the critical wavenumber. They indicate that the long wavelength QEA waves (i.e. for lower values of $k$, i.e.  $k<k_c$)
are modulationally unstable, and the corresponding solution of the NLS equation gives rise to the bright solitons.
On the otherhand,
the short wavelength QEA waves (i.e. for higher values of $k$, i.e.  $k>k_c$) becomes modulationally stable, and the corresponding solution of the NLS equation gives rise to the dark solitons.  It is also clear from figures  \ref{f6} and \ref{f7} that  the critical wavenumber $k_c$ decreases (increases) as we increase the spectral index $\kappa$
($\alpha$).
We are interested in the solution corresponding to the bright solitons  (i.e.  $P/Q>0$)  of  the NLS equation, (\ref{NLS}),
which is given by  \cite{ref57,ref58}:
\begin{eqnarray}
a(\xi,\tau) = a_0\left(\sqrt{\frac{P}{Q}}\right)\exp{[iP\tau(\tau)]},
\label{sol}
\end{eqnarray}
where $a_0 (x,\tau)= \sqrt{2}[{(4+i16P\tau)}/{(1+16P^2\tau^2+4\xi^2)}-1]$.
The solution (\ref{sol}) predicts the concentration of the QEA wave in a small region due to the nonlinear properties of the plasma, and it is able to concentrate a significant amount of the wave energy into a relatively small area in space \cite{ref57}.  We have graphically shown  the time dependent  bright (envelope)  solitons, i. e.  the variation of
$a*a=|a|^2$ with the position ($\xi$) and time ($\tau$). This is displayed in figure \ref{f8} which shows how the QEA envelope solitonic profile evolve with time. This surface plot indicates that the QEA waves are  localized in both $\xi$ and $\tau$  axes.  This feature means that the nonlinear QEA waves can also concentrate the energy of the plasma system in a small region \cite{ref59}.  The width of the localized  structures get flattened along the $\tau$ axis. On the other hand, the stationary envelope solitonic profiles for different values of $\kappa$ and $\alpha$ are shown in figures \ref{f9} and \ref{f10}, respectively.
It is obvious from figures \ref{f9}  and \ref{f10} that as we increase the value of $\kappa$  or $\alpha$, the amplitude of the  QEA envelope solitons increases, but their width remains unchanged.
\section{Discussion}
We have considered a three-component degenerate quantum plasma (DQP) system  containing cold quantum electron fluid \cite{ref12,ref13}, inertialess, superthermal \cite{MH1990,Mace-1991} electrons, and uniformly distributed stationary ions \cite{MS2002} to identify the effects of  suprathermality \cite{ref47} of hot electron component,  the degenerate cold electron pressure, cold electron exchange correlation potential, and   Bohm potential of cold electron component on the linear and nonlinear properties of the QEA waves.  We have derived the NLS equation by the reductive perturbation method, and have obtained its solitonic solution to find the basic features of the QEA envelope solitons. The results, which have been found from this theoretical investigation, can be pinpointed as follows:
\begin{enumerate}
\item{The quantum effect due to the degenerate electron pressure of the cold electron species dominates
over that due to the particle exchange-correlation potential or the Bohm potential on the dispersion properties of  the
 long wavelength QEA waves. However, as the wavelength of the  QEA waves is decreased,
the effect of the Bohm potential overtakes that of the exchange-correlation potential.}

\item{It is found that for a long wavelength limit (which corresponds to a very low $k$-value range) the angular frequency  $\omega$ linearly increases with $k$, and for a short wavelength limit (which corresponds to a very high
$k$-value range)  it is independent of $k$  (saturated region). This is usual dispersion properties of any kind of acoustic-type of waves. It is also observed that as we  increase $\kappa$ ($\alpha$), the $\omega$ vs.  $k$  curve is shifted up (down) to  $\omega-$ axis, and the saturation region is reached  for higher values of  $\kappa$ and
$\alpha$.}

\item{The long wavelength QEA waves (satisfying $k<k_c$)  are modulationally unstable, and the corresponding solution of the NLS equation gives rise to  the bright solitons, where $k_c$ is the minimum value of $k$ above which the QEA waves are modulationally stable.  On the otherhand,
the short wavelength QEA waves (satifying $k>k_c$) becomes modulationally stable,
the corresponding solution of the NLS equation gives rise to  the dark  solitons.  It is  observed that $k_c$ is decreased as the spectral index $\kappa$ is increased, and that  it is independent of $\alpha$.}

\item{It is seen that as  $\kappa$ ($\alpha$) increases,  the group velocity $v_0$  increases for lower (higher) values of $\kappa$ ($\alpha$),  and becomes very sharp at the low value ranges of $\kappa$ and $\alpha$.}

\item{It is observed that the QEA waves are  localized (as bright envelope solitons) in both $\xi$ and $\tau$  axes, and that  as the value of $\kappa$  or $\alpha$ is increased, the amplitude of the  QEA envelope solitons increases, but their width remains unchanged.  This feature means that the nonlinear waves can concentrate the energy of the plasma system in its small region \cite{ref59}. }
\end{enumerate}

To conclude, we stress that  our present investigation on the QEA waves and associated instability and nonlinear structures in a DQP (containing cold quantum electron fluid  \cite{ref12,ref13} with Fermi energy $E_F$, inertialess, superthermal \cite{MH1990,Mace-1991} electron component. and uniformly distributed stationary ions \cite{MS2002})  is expected to help us to  understand the localized low-frequency  electrostatic disturbances in laboratory solid density plasma produced by irradiating metals by intense laser,  semiconductor devices, microelectronics, carbon nanotubes, etc. \cite{Jung2001,Ang2003,Killian2006,Shah2012}.  We also suggest to perform a laboratory solid density plasma experiment based on the parameters used in our numerical analysis, which may be able to identify the basic features of linear and nonlinear QEA waves predicted in our present investigation.
\\\\
{\large {\bf Appendix}}\\\\
The notations $f_0$, $f_1$, $f_2$, $g_1$, and $g_2$ appearing in  (\ref{NLS1}) and (\ref{NLS2})
are listed as follows:
\begin{eqnarray}
&&f_0 = \left[k(X_2-\omega a_1a_2)-\nu\Omega\right]X_1+\left[2\nu V_0-Y_2\right]Y_1,\nonumber \\
&&f_1=v_0a_6-a_8, \nonumber \\
&&f_2=v_0a_8-(\sigma+\beta+3k^2\mu)a_6+3k\mu a_1,\nonumber \\
&&g_1=-k[a_1(a_{17}+a_{24})+a_2(a_{16}+a_{23})],\nonumber \\
&&g_2=k[-a_2(a_{17}+a_{24})+a_1g_{21}+k^2\mu a_1g_{22}],\nonumber
\end{eqnarray}
where
\begin{eqnarray}
&&X_1=[-k^2\alpha+(4k^2+\delta)\Omega]^{-1}, \nonumber\\
&&X_2=k(X_\sigma a_1^2-a_2^2)/2, \nonumber \\
&&Y_1=[(\sigma+\beta-v_0^2)\delta+\alpha]^{-1}, \nonumber\\
&&Y_2=X_\sigma a_1^2 +a_2^2+2v_0a_1a_2, \nonumber\\
&&X_\sigma=(\sigma+2\beta)/3+k^2\mu, \nonumber\\
&&\Omega=\omega^2-k^2b_1, \nonumber\\
&&V_0=\sigma+\beta-v_0^2, \nonumber\\
&&g_{21}=(\sigma+2(a_{16}+a_{23})/3, \nonumber\\
&&g_{22}=7a_{16}+a_{23}, \nonumber\\
&&b_1=\sigma+\beta+4k^2\mu^2, \nonumber\\
&&a_3=v_0a_1-a_2, \nonumber\\
&&a_4=v_0a_2+\alpha-a_1(\sigma+\beta+3k^2\mu^2), \nonumber\\
&&a_6 = (\omega a_3+ka_4)/(\omega^2-k^2b_0), \nonumber\\
&&a_8=(\omega_6-a_3)/k, \nonumber\\
&&a_9=-2a_1a_2, \nonumber\\
&&a_{10}=2\beta a_1^2/3+k^2\mu a_1^2, \nonumber\\
&&a_{11}=-k^2\alpha/(\omega^2-k^2b_1), \nonumber\\
&&a_{12}=(\omega a_9-ka_{10})/(\omega^2-k^2b_1), \nonumber\\
&&a_{13}=\omega a_{11}/k, \nonumber\\
&&a_{14}=\omega(a_{12}-a_1^2)/k, \nonumber\\
&&a_{15}=-(a_{12}+\nu/(4k^2+\delta+a_{11}), \nonumber\\
&&a_{16}=a_{11}a_{15}+a_{12}, \nonumber\\
&&a_{17}=a_{13}a_{15}+a_{14}, \nonumber\\
&&a_{18}=2a_1a_2, \nonumber\\
&&a_{19}=-a_2^2+(\sigma+2\beta+k^2\mu)/3a_1^2, \nonumber\\
&&a_{20}=\alpha/V_0, \nonumber\\
&&a_{21}=(a_{19}-v_0a_{18})/V_0, \nonumber\\
&&a_{22}=(2\nu-a_{21})/(\delta+a_{20}), \nonumber\\
&&a_{23}=a_{20}a_{22}+a_{21}, \nonumber\\
&&a_{24}=a_{24}=v_0a_{23}-a_{18}. \nonumber
\end{eqnarray}

\end{document}